\documentclass[twocolumn,showpacs,preprintnumbers,amsmath,amssymb,prd,superscriptaddress]{revtex4-1}
\usepackage[section]{placeins}

\usepackage{graphicx}
\usepackage{dcolumn}
\usepackage{bm}
\usepackage{multirow}
\usepackage{array}
\usepackage{tabularx}
\usepackage{xcolor}
\usepackage{lipsum}
\usepackage{enumitem}
\usepackage{setspace}
\usepackage[compact]{titlesec}

\usepackage{amsmath}
\usepackage[mathlines]{lineno}

\newcommand{\etah}{\eta_{\Lambda(\overline{\Lambda})}}
\newcommand{\etaj}{\eta_{jet}}
\newcommand{\pt}{p_T}
\newcommand{\pth}{p_{T,{\Lambda(\overline{\Lambda})}}}
\newcommand{\GeV}{\mathrm{GeV}}
\renewcommand{\arraystretch}{1}
\newcommand{\DLL}{D_{LL}}
\newcommand{\DTT}{D_{TT}}
\newcommand*{\ora}{\overrightarrow}

\titlespacing{\section}{0pt}{5pt}{5pt}
\titlespacing{\subsection}{0pt}{5pt}{5pt}
\titlespacing{\subsubsection}{0pt}{10pt}{5pt}
\let\oldequation\equation
\let\oldendequation\endequation

\renewenvironment{equation}
  {\linenomathNonumbers\oldequation}
  {\oldendequation\endlinenomath}

\let\oldalign\align
\let\oldendalign\endalign

\renewenvironment{align}
  {\linenomathNonumbers\oldalign}
  {\oldendalign\endlinenomath}
\begin{document}
\bibliographystyle{unsrt}
\title{~\\ Longitudinal and transverse spin transfer to $\Lambda$ and $\overline{\Lambda}$ hyperons in polarized $p$+$p$ collisions at $\sqrt{s} = 200$ GeV} 

\affiliation{Abilene Christian University, Abilene, Texas   79699}
\affiliation{AGH University of Krakow, FPACS, Cracow 30-059, Poland}
\affiliation{Argonne National Laboratory, Argonne, Illinois 60439}
\affiliation{American University in Cairo, New Cairo 11835, Egypt}
\affiliation{Ball State University, Muncie, Indiana, 47306}
\affiliation{Brookhaven National Laboratory, Upton, New York 11973}
\affiliation{University of Calabria \& INFN-Cosenza, Rende 87036, Italy}
\affiliation{University of California, Berkeley, California 94720}
\affiliation{University of California, Davis, California 95616}
\affiliation{University of California, Los Angeles, California 90095}
\affiliation{University of California, Riverside, California 92521}
\affiliation{Central China Normal University, Wuhan, Hubei 430079 }
\affiliation{University of Illinois at Chicago, Chicago, Illinois 60607}
\affiliation{Creighton University, Omaha, Nebraska 68178}
\affiliation{Czech Technical University in Prague, FNSPE, Prague 115 19, Czech Republic}
\affiliation{Technische Universit\"at Darmstadt, Darmstadt 64289, Germany}
\affiliation{National Institute of Technology Durgapur, Durgapur - 713209, India}
\affiliation{ELTE E\"otv\"os Lor\'and University, Budapest, Hungary H-1117}
\affiliation{Frankfurt Institute for Advanced Studies FIAS, Frankfurt 60438, Germany}
\affiliation{Fudan University, Shanghai, 200433 }
\affiliation{University of Heidelberg, Heidelberg 69120, Germany }
\affiliation{University of Houston, Houston, Texas 77204}
\affiliation{Huzhou University, Huzhou, Zhejiang  313000}
\affiliation{Indian Institute of Science Education and Research (IISER), Berhampur 760010 , India}
\affiliation{Indian Institute of Science Education and Research (IISER) Tirupati, Tirupati 517507, India}
\affiliation{Indian Institute Technology, Patna, Bihar 801106, India}
\affiliation{Indiana University, Bloomington, Indiana 47408}
\affiliation{Institute of Modern Physics, Chinese Academy of Sciences, Lanzhou, Gansu 730000 }
\affiliation{University of Jammu, Jammu 180001, India}
\affiliation{Kent State University, Kent, Ohio 44242}
\affiliation{University of Kentucky, Lexington, Kentucky 40506-0055}
\affiliation{Lawrence Berkeley National Laboratory, Berkeley, California 94720}
\affiliation{Lehigh University, Bethlehem, Pennsylvania 18015}
\affiliation{Max-Planck-Institut f\"ur Physik, Munich 80805, Germany}
\affiliation{Michigan State University, East Lansing, Michigan 48824}
\affiliation{National Institute of Science Education and Research, HBNI, Jatni 752050, India}
\affiliation{National Cheng Kung University, Tainan 70101 }
\affiliation{Nuclear Physics Institute of the CAS, Rez 250 68, Czech Republic}
\affiliation{The Ohio State University, Columbus, Ohio 43210}
\affiliation{Institute of Nuclear Physics PAN, Cracow 31-342, Poland}
\affiliation{Panjab University, Chandigarh 160014, India}
\affiliation{Purdue University, West Lafayette, Indiana 47907}
\affiliation{Rice University, Houston, Texas 77251}
\affiliation{Rutgers University, Piscataway, New Jersey 08854}
\affiliation{Universidade de S\~ao Paulo, S\~ao Paulo, Brazil 05314-970}
\affiliation{University of Science and Technology of China, Hefei, Anhui 230026}
\affiliation{South China Normal University, Guangzhou, Guangdong 510631}
\affiliation{Sejong University, Seoul, 05006, South Korea}
\affiliation{Shandong University, Qingdao, Shandong 266237}
\affiliation{Shanghai Institute of Applied Physics, Chinese Academy of Sciences, Shanghai 201800}
\affiliation{Southern Connecticut State University, New Haven, Connecticut 06515}
\affiliation{State University of New York, Stony Brook, New York 11794}
\affiliation{Instituto de Alta Investigaci\'on, Universidad de Tarapac\'a, Arica 1000000, Chile}
\affiliation{Temple University, Philadelphia, Pennsylvania 19122}
\affiliation{Texas A\&M University, College Station, Texas 77843}
\affiliation{University of Texas, Austin, Texas 78712}
\affiliation{Tsinghua University, Beijing 100084}
\affiliation{University of Tsukuba, Tsukuba, Ibaraki 305-8571, Japan}
\affiliation{University of Chinese Academy of Sciences, Beijing, 101408}
\affiliation{United States Naval Academy, Annapolis, Maryland 21402}
\affiliation{Valparaiso University, Valparaiso, Indiana 46383}
\affiliation{Variable Energy Cyclotron Centre, Kolkata 700064, India}
\affiliation{Warsaw University of Technology, Warsaw 00-661, Poland}
\affiliation{Wayne State University, Detroit, Michigan 48201}
\affiliation{Yale University, New Haven, Connecticut 06520}

\author{M.~I.~Abdulhamid}\affiliation{American University in Cairo, New Cairo 11835, Egypt}
\author{B.~E.~Aboona}\affiliation{Texas A\&M University, College Station, Texas 77843}
\author{J.~Adam}\affiliation{Czech Technical University in Prague, FNSPE, Prague 115 19, Czech Republic}
\author{L.~Adamczyk}\affiliation{AGH University of Krakow, FPACS, Cracow 30-059, Poland}
\author{J.~R.~Adams}\affiliation{The Ohio State University, Columbus, Ohio 43210}
\author{I.~Aggarwal}\affiliation{Panjab University, Chandigarh 160014, India}
\author{M.~M.~Aggarwal}\affiliation{Panjab University, Chandigarh 160014, India}
\author{Z.~Ahammed}\affiliation{Variable Energy Cyclotron Centre, Kolkata 700064, India}
\author{D.~M.~Anderson}\affiliation{Texas A\&M University, College Station, Texas 77843}
\author{E.~C.~Aschenauer}\affiliation{Brookhaven National Laboratory, Upton, New York 11973}
% ...

\collaboration{STAR Collaboration}\noaffiliation
\date{\today}

\begin{abstract}
The longitudinal and transverse spin transfers to $\Lambda$ ($\overline{\Lambda}$) hyperons in polarized proton-proton collisions are expected to be sensitive to the helicity and transversity distributions, respectively, of (anti-)strange quarks in the proton, and to the corresponding polarized fragmentation functions. We report improved measurements of the longitudinal spin transfer coefficient, $\DLL$, and the transverse spin transfer coefficient, $\DTT$, to $\Lambda$ and $\overline{\Lambda}$ in polarized proton-proton collisions at $\sqrt{s}$ = 200 GeV by the STAR experiment at RHIC. The data set includes longitudinally polarized proton-proton collisions with an integrated luminosity of 52 pb$^{-1}$, and transversely polarized proton-proton collisions with a similar integrated luminosity. Both data sets have about twice the statistics of previous results and cover a kinematic range of $|\etah|$ $<$ 1.2 and transverse momentum $\pth$ up to 8 $\GeV/c$. We also report the first measurements of the hyperon spin transfer coefficients $\DLL$ and $\DTT$ as a function of the fractional jet momentum $z$ carried by the hyperon, which can provide more direct constraints on the polarized fragmentation functions.  
\end{abstract}

\maketitle

\section{Introduction}

The spin structure of hadrons, in particular the nucleon, remains a fundamental question in the field of Quantum Chromodynamics (QCD). 
Tremendous progress has been made in recent years on the helicity distributions of the nucleon, including the gluon spin contribution and light sea quark spin contributions,  
with strange quark helicity distributions less constrained\,\cite{Nocera:2014gqa,Ethier:2017zbq,DeFlorian:2019xxt,Alexandrou:2020sml}. 
For the transversity distributions,  good progress has also been made on the valence quark distributions through semi-inclusive deep inelastic scattering (DIS) and proton-proton collisions, with still poor knowledge on sea quark transversity\,\cite{Kang:2015msa,Perdekamp:2015vwa,Radici:2018iag,Anselmino:2020nrk}.
Due to their self spin-analyzing parity-violating decay\,\cite{Lee:1957qs,Crawford:1957zzb,Cronin:1963zb},
$\Lambda$ and $\overline{\Lambda}$ polarizations have been studied extensively in DIS and proton-proton collisions, which provide unique opportunities to study nucleon spin structure and spin effects in the hadronization process\,\cite{Bunce:1976yb,Artru:1990wq,Collins:1993kq}. 
Experiments in which proton beam polarization is transferred to outgoing $\Lambda$ polarization (usually referred to as ``spin transfer'') provide connections to the polarized parton densities of the proton and the polarized fragmentation functions of the hyperon. 
In particular, as the (anti-)strange quark plays a dominant role in the $\Lambda\,(\overline{\Lambda})$ hyperon's spin content, measurements of the spin transfer coefficient to $\Lambda\,(\overline{\Lambda})$ hyperons provide a way to gain insights into the polarized distribution of (anti-) strange quarks in the nucleon\,\cite{Artru:1990wq,Collins:1993kq,Lu:1995np,Ellis:1995fc,Jaffe:1996wp,Ma:2000uu,Ellis:2002zv,Zhou:2009mx,Xu:2005ru,Chen:2007tm,Liu:2019xcf}.

The longitudinal spin transfer to $\Lambda$ ($\overline{\Lambda}$) hyperons in lepton-nucleon\,\cite{Lu:1995np,Ellis:1995fc,Jaffe:1996wp,Ma:2000uu,Ellis:2002zv,Zhou:2009mx} and proton-proton collisions\,\cite{deFlorian:1998ba,Boros:2000ya,Ma:2001na,Xu:2002hz,Xu:2004es,Xu:2005ru,Chen:2007tm,Liu:2019xcf} provides sensitivity to the helicity distribution of (anti-)strange quarks through polarized fragmentation functions.
Similarly, with a transversely polarized proton beam, the transverse spin transfer to $\Lambda$ ($\overline{\Lambda}$) in lepton-nucleon and proton-proton collisions provides a natural connection to the transversity distribution of \mbox{(anti-)} strange quarks through transversely polarized fragmentation functions\,\cite{Artru:1990wq,Collins:1993kq,deFlorian:1998am,Xu:2004es,Xu:2005ru,COMPASS:2021bws,Kang:2021kpt}.
The transversity distribution remains less understood than the helicity distribution due to its chiral-odd nature\,\cite{Barone:2001sp,Perdekamp:2015vwa}, and currently, almost no experimental data have provided any constraints on the strange quark transversity\,\cite{Kang:2015msa,Radici:2018iag}.  
On the other hand, the polarized fragmentation functions provide key information about the spin content of hyperons, which cannot be probed directly through scattering experiments with hyperons. 
Recently, it has been shown that measuring the spin transfer coefficients as a function of the jet momentum fraction $z$ carried by the $\Lambda$ ($\overline{\Lambda}$) hyperon can directly probe the polarized jet fragmentation functions of the  $\Lambda$ ($\overline{\Lambda}$)\,\cite{Kang:2020xyq}.
A number of measurements of $\Lambda$ ($\overline{\Lambda}$) hyperon spin transfer coefficients have been made in past years in polarized lepton-nucleon DIS experiments\,\cite{COMPASS:2009nhs,COMPASS:2021bws,Adamsetal._2000_LambdaBar,HERMES_2006_LongitudinalSpin}, and in polarized proton-proton collisions\,\cite{Bravar:1997fb,Abelev:2009xg,Adam:2018dtt,Adam:2018kzl}. 
New, high precision measurements of hyperon spin transfer coefficients are needed to gain further knowledge about the polarized parton distributions and the polarized fragmentation functions. 
The high-luminosity proton-proton ($pp$) collisions available at the Relativistic Heavy Ion Collider (RHIC), with both beams polarized, provide a unique opportunity for such measurements.

In this paper, we report improved measurements of the longitudinal spin transfer coefficient $\DLL$ and the transverse spin transfer coefficient $\DTT$ of  $\Lambda$ and $\overline{\Lambda}$ hyperons as a function of the hyperon transverse momentum $\pth$ in polarized $pp$  collisions at $\sqrt s=200$\,GeV by the Solenoidal Tracker At RHIC (STAR) experiment. 
About twice the hyperon statistics of previous measurements\,\cite{Adam:2018dtt,Adam:2018kzl} were used for both coefficients. In addition, we report the first measurements of the spin transfer coefficients $\DLL$ and $\DTT$ as a function of the fractional jet momentum $z$ carried by the hyperon, which provide a direct probe of the polarized fragmentation functions.

The spin transfer coefficients of hyperons $\DLL$ and $\DTT$ in $pp$ collisions are defined as follows: \\
(i): The longitudinal spin transfer coefficient, $\DLL$,  in proton-proton collisions is defined as:
\begin{align}
\centering
\DLL &\equiv \frac
{d\sigma^{\left[p^{+(-)}p \to  \Lambda ^{+(-)} X\right]}-d\sigma^{\left[p^{+(-)}p \to  \Lambda ^{-(+)}X\right]}}
{d\sigma^{\left[p^{+(-)}p \to  \Lambda ^{+(-)} X\right]}+d\sigma^{\left[p^{+(-)}p \to  \Lambda ^{-(+)}X\right]}}\notag{} \\
  & = \frac{d\Delta\sigma^{\Lambda}}{d\sigma^{\Lambda}},\label{eq:dllDef}
\end{align}

where the superscripts $+$ or $-$ denote the helicity of the proton beam or the $\Lambda$ hyperon, and $\Delta\sigma^{\Lambda}$ is the longitudinally polarized cross section.
Within a factorized  framework, the polarized cross section can be described as the convolution of the parton helicity distributions of the proton, the polarized cross section of partonic scattering, and the longitudinally polarized fragmentation function of hyperon.
Thus, measurements of $\DLL$ to $\Lambda$ and $\overline{\Lambda}$ can provide insights into the strange quark and anti-quark helicity distributions and the longitudinally polarized fragmentation functions\,\cite{deFlorian:1998ba,Boros:2000ya,Ma:2001na,Xu:2002hz,Xu:2005ru,Chen:2007tm}. \\
(ii): The transverse spin transfer coefficient, $\DTT$, in proton-proton collisions is defined as: 
\begin{align}
  \centering
  \DTT&\equiv \frac{d\sigma^{{\left[p^{\uparrow(\downarrow)}p \rightarrow \Lambda^{\uparrow(\downarrow)}X\right]}}-d\sigma^{{\left[p^{\uparrow(\downarrow)}p \rightarrow \Lambda^{\downarrow(\uparrow)}X\right]}}} {d\sigma^{{\left[p^{\uparrow(\downarrow)}p \rightarrow \Lambda^{\uparrow(\downarrow)}X\right]}}+ d\sigma^{{\left[p^{\uparrow(\downarrow)}p \rightarrow \Lambda^{\downarrow(\uparrow)}X\right]}}} \notag{} \\
  &= \frac{d\delta\sigma^{\Lambda}}{d\sigma^{\Lambda}}, \label{eq:dttDef}
\end{align}
where $\uparrow$\,($\downarrow$) denotes the upward\,(downward) transverse polarization direction of the particles and $\delta\sigma^{\Lambda}$ is the transversely polarized cross section.
Similarly, $\delta\sigma^{\Lambda}$ can be written as the convolution of the quark transversity of the proton, the polarized cross section of partonic scattering, and the polarized fragmentation function\,\cite{deFlorian:1998am} of hyperon. Thus, the measurements of $\DTT$ provide natural connections to quark transversity and the polarized fragmentation functions\,\cite{deFlorian:1998am,Xu:2004es,Xu:2005ru}.

\begin{figure}[htbp]
\centering
\includegraphics[width=1\linewidth]{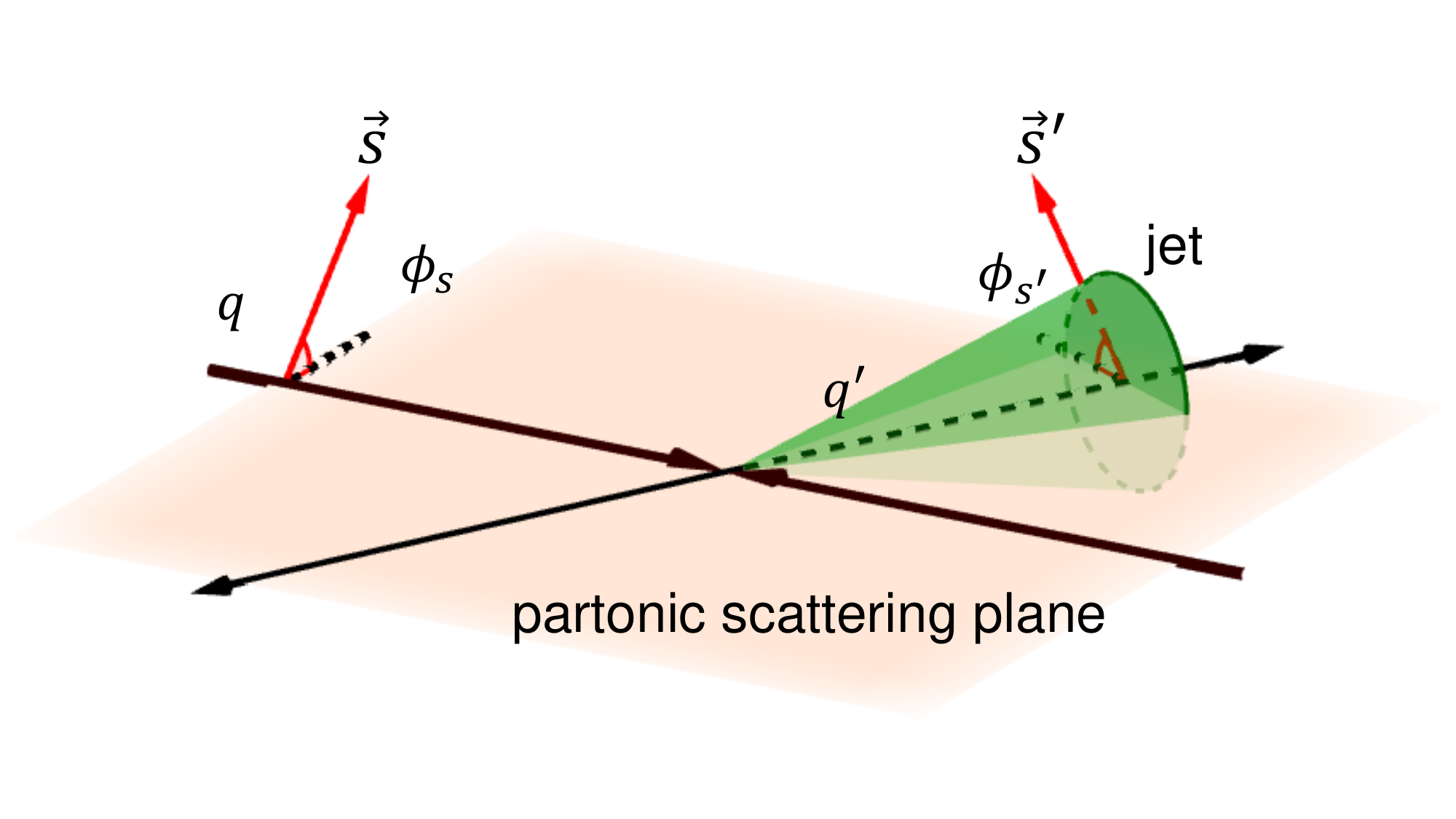}
\caption{Illustration of spin vectors for initial $(\Vec{S})$ and final $(\Vec{S^{'}})$ state quarks during partonic scattering in transversely polarized proton-proton collisions. The corresponding azimuth angles of the spin vectors, $\phi_{S}$ and $\phi_{S^{'}}$, are equal\,\cite{Collins:1993kq}.}
\label{fig:DTT-carton}
\end{figure}

The polarization of $\Lambda\,(\overline{\Lambda})$ hyperons, $P_{\Lambda\,(\overline{\Lambda})}$, can be determined experimentally from the angular distribution of their decay daughters via the weak decay channel $\Lambda\rightarrow p\pi^{-}\,(\overline{\Lambda}\rightarrow \overline{p}\pi^{+})$\,\cite{Lee:1957qs,Crawford:1957zzb,Cronin:1963zb},
\begin{equation}
  \frac
  {dN}{\,d\cos{\theta^{*}}\,}
  \propto
  A
  \left(1+\alpha_{\Lambda\,(\overline{\Lambda})}P_{\Lambda\,(\overline{\Lambda})}\cos{\theta^{*}}\right),
  \label{eq:distFinalState}
\end{equation}
where $A$ is the detector acceptance (varies with $\theta^{*}$ and other observables), $\alpha_{\Lambda\,(\overline{\Lambda})}$ is the weak decay parameter, and $\theta^{*}$ is the angle between the $\Lambda\,(\overline{\Lambda})$ polarization direction and the daughter (anti-)proton momentum in the $\Lambda\,(\overline{\Lambda})$ rest frame. 
For the $\DLL$ measurements, the  polarization direction is taken to be along the moving direction of the $\Lambda\,(\overline{\Lambda})$ in the $pp$ center-of-mass frame (also the lab frame).
But for the $\DTT$ measurements, the transverse polarization direction of the outgoing fragmenting parton is used to obtain $\theta^{*}$\,\cite{Adam:2018dtt}.
Because there is a rotation along the normal direction to the scattering plane between the spin vectors of the initial and final state quarks\,\cite{Collins:1993kq} (as shown in Fig.\ref{fig:DTT-carton}), the momentum direction of the outgoing parton is required. The reconstructed jet axis is used as a substitute for the direction of the outgoing fragmenting quark\,\cite{Adam:2018dtt}.

\section {Experimental details and data analysis}

\subsection{Data sample and event selection}

The data were collected with proton-proton collisions at $\sqrt s=200$ GeV at RHIC with the STAR detector in the year 2015, corresponding to a  sampled luminosity of 52 $\mathrm{pb^{-1}}$ for longitudinally polarized $pp$ collisions and a similar number for transversely polarized $pp$ collisions. The proton polarizations were measured for each beam and each beam fill using Coulomb-Nuclear Interference proton-carbon polarimeters\,\cite{pCPol} calibrated using a polarized atomic hydrogen gas-jet target. 
The average polarizations of the two beams were $56\%$ and $51\%$ for longitudinally polarized beams, and were $57\%$ and $57\%$ for transversely polarized beams.
\begin{table*}[htbp]
\scriptsize
\centering
\caption{
Selection cuts for $\Lambda(\overline{\Lambda})$ reconstruction: the upper part is for candidates with daughter $\pi^{-}(\pi^{+})$ matched to a TOF hit, and the lower part is for candidates without a TOF match. 
Here, “DCA” denotes ``distance of closest approach'', ``PV'' denotes  ``primary vertex'',  
$\protect\ora{r}$ denotes the vector from the primary vertex to the decay vertex of $\Lambda$ or $\overline{\Lambda}$
and $\protect\ora{p}$ denotes the momentum vector of $\Lambda$ or $\overline{\Lambda}$.}
\renewcommand{\arraystretch}{1.3}
\resizebox{0.85\linewidth}{!}{
\begin{tabular}{ c | c | c | c | c | c | c }
            \hline
		\multicolumn{7}{c}{$\pi^{\pm}$ matches a TOF hit} \\	\hline
		$\pth$ ($\GeV/c$) & $ < 2$  & $2- 3$ & $3- 4$ & $4-5$ & $5 - 6$ & $ > 6$ \\
		\hline
		DCA of $p(\overline{p})$ to PV  & $>$ 0.2 cm & $>$ 0.15 cm & $>$ 0.05 cm & $>$ 0.005 cm & $>$ 0.005 cm & $>$ 0.005 cm\\
		\hline
		DCA of $\pi^{-}(\pi^+)$ to PV & $>$ 0.6 cm & $>$ 0.55 cm & $>$ 0.5 cm & $>$ 0.5 cm & $>$ 0.5 cm & $>$ 0.5 cm\\
		\hline
		DCA of $p\pi^-$ ($\overline{p}\pi^+$) & $<$ 0.75 cm & $<$ 0.65 cm & $<$ 0.6 cm & $<$ 0.5 cm & $<$ 0.45 cm & $<$ 0.45 cm\\
		\hline
		DCA of $\Lambda$($\overline{\Lambda}$) to PV & $<$ 1 cm & $<$ 1 cm & $<$ 1 cm & $<$ 1 cm & $<$ 1 cm & $<$ 1 cm\\
		\hline
		Decay Length & $>$ 3 cm & $>$ 3.5 cm & $>$ 3.5 cm & $>$ 4 cm & $>$ 4.5 cm & $>$ 4.5 cm\\
		\hline
		$\mathrm{cos}(\overrightarrow{r},\overrightarrow{p})$ & $>$ 0.995 & $>$ 0.995 & $>$ 0.995 & $>$ 0.995 & $>$ 0.995 & $>$ 0.995\\
		\hline\hline
		\multicolumn{7}{c}{$\pi^{\pm}$ does not match a TOF hit} \\	\hline

            $\pth$ ($\GeV/c$) & $ < 2$  & $2- 3$ & $3- 4$ & $4- 5$ & $5- 6$ & $> 6$ \\
		\hline
		DCA of $p(\overline{p})$ to PV  & $>$ 0.45 cm & $>$ 0.3 cm & $>$ 0.25 cm & $>$ 0.2 cm & $>$ 0.15 cm & $>$ 0.15 cm\\
		\hline
		DCA of $\pi^{-}(\pi^+)$ to PV & $>$ 0.65 cm & $>$ 0.6 cm & $>$ 0.55 cm & $>$ 0.55 cm & $>$ 0.55 cm & $>$ 0.5 cm\\
		\hline
		DCA of $p\pi^-$ ($\overline{p}\pi^+$) & $<$ 0.7 cm & $<$ 0.6 cm & $<$ 0.55 cm & $<$ 0.5 cm & $<$ 0.45 cm & $<$ 0.45 cm\\
		\hline
		DCA of $\Lambda$($\overline{\Lambda}$) to PV & $<$ 0.55 cm & $<$ 0.55 cm & $<$ 0.6 cm & $<$ 0.6 cm & $<$ 0.6 cm & $<$ 0.6 cm\\
		\hline
		Decay Length & $>$ 7 cm & $>$ 7 cm & $>$ 7 cm & $>$ 8.5 cm & $>$ 10 cm & $>$ 10.5 cm\\
		\hline
		$\mathrm{cos}(\overrightarrow{r},\overrightarrow{p})$ & $>$ 0.995 & $>$ 0.995 & $>$ 0.995 & $>$ 0.995 & $>$ 0.995 & $>$ 0.995\\
		\hline
	\end{tabular}
 }
 
	\label{table:L_cut}
\end{table*}

The subsystems of the STAR detector\,\cite{STAR_det} used in these measurements are the Time Projection Chamber (TPC)\,\cite{TPC}, the Barrel Electromagnetic Calorimeter (BEMC)\,\cite{BEMC}, the Endcap Electromagnetic Calorimeter (EEMC)\,\cite{EEMC}, the Time of Flight (TOF) detector\,\cite{TOF}, the Vertex Position Detectors (VPD)\,\cite{VPD}, and the Zero Degree Calorimeters (ZDC)\,\cite{ZDC}.
The TPC covers the pseudorapidity range $|\eta| \lesssim 1.3$ and $2\pi$ in azimuthal direction. It measures the trajectories of the charged particles in a 0.5 T 
magnetic field.
Particle identification is made through the ionization energy loss ($dE/dx$) of a charged particle in the TPC gas. 
The BEMC and the EEMC cover $|\eta| < 1.0$ and $1.086 < \eta < 2.0$, respectively, with full azimuthal angle coverage.
The TOF covers $|\eta|<0.9$ and $2\pi$ in azimuthal angle. It provides additional particle identification by measuring the flight time of charged particles.

The VPD and ZDC, which cover pseudorapidity $4.2<|\eta|<5.2$ and $|\eta|>6.6$, respectively, are used to monitor the luminosity ratios for the different polarization states of the colliding beams.
The jet-patch (JP) triggers are used in the event selection, which require the transverse electromagnetic energy, $E_T$, in a region $\Delta \eta \times \Delta \phi = 1.0\times 1.0$ in the BEMC and EEMC to exceed a given threshold. In 2015, the thresholds were $E_T=5.4$ $\GeV$ (JP1, prescaled) and $E_T=7.3$ $\GeV$ (JP2).
In addition, the $z$ component of the primary vertex (PV) determined with TPC tracks for each event is required to be within 90 cm of the center of the TPC along the beam line to ensure uniform acceptance.

\subsection{$\Lambda\,(\overline{\Lambda})$ and jet reconstruction}

Similar to previous published measurements\,\cite{Abelev:2009xg, Adam:2018dtt, Adam:2018kzl}, in this analysis the $\Lambda$ ($\overline{\Lambda}$) is reconstructed via its decay channel $\Lambda\to p\pi^-$ ($\overline{\Lambda} \to \overline{p}\pi^+$), corresponding to a branching ratio of about 64.1\%\,\cite{PDG}. Daughter candidates are identified based on their charge sign and energy loss inside the TPC. Two daughter candidates are then paired, and a set of selection criteria based on decay topology is applied to select the hyperon candidates, with the residual background at an acceptable level (below or around 10\%).
The selection criteria vary with hyperon $\pth$. 
Due to geometric acceptance and detector inefficiencies, only about 50\% of the decay pions could be matched to a TOF hit.
The signal is much cleaner when the hyperon daughter pion track matches a TOF hit, as the response time of TOF is much shorter than that of the TPC, and the TOF matching helps to remove pile-up tracks. 
Correspondingly, the selection criteria are divided into two groups based on whether the daughter pion track matches a TOF hit or not. 
Tighter cuts are applied to the sample without TOF matching to reduce the random background. 
The final fraction of $\Lambda$ and $\overline{\Lambda}$ candidates with pion matched to TOF is about 70\% after all the cuts. 
The selection criteria are summarized separately in Table~\ref{table:L_cut} for these two cases separately.

\begin{figure*}[htbp]
\centering
\includegraphics[width=0.9\linewidth]{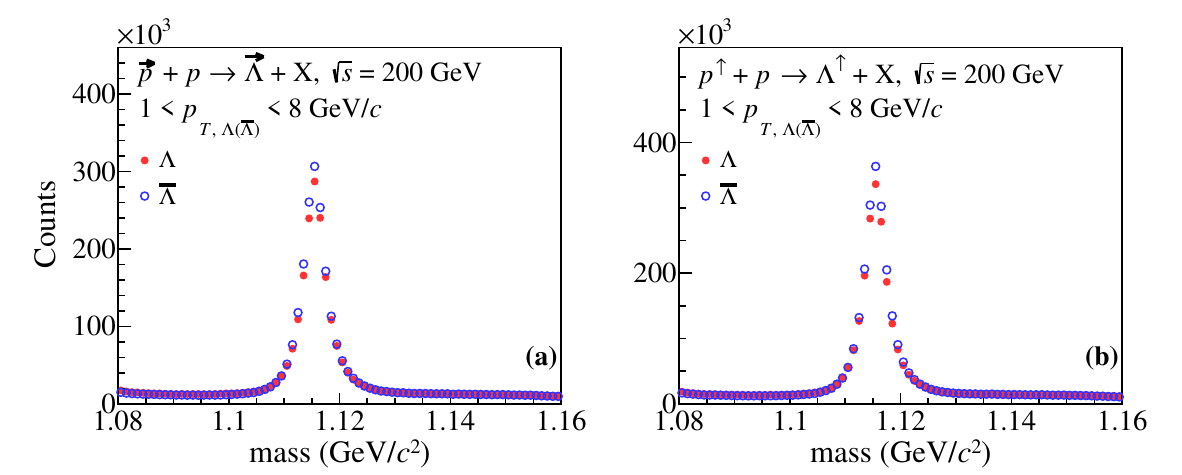}
\caption{Invariant mass spectra of $\Lambda$ (closed circles) and $\overline{\Lambda}$ (open circles) candidates with $1< \pth < 8$ GeV$/c$  from (a) longitudinally and (b) transversely polarized proton-proton collisions at $\sqrt{s} = 200$ GeV.}
\label{fig:im_L}
\end{figure*}

In this analysis, the spin transfer coefficients are measured for the hyperons in jets, which means the hyperons are among the fragments of a hard scattered parton.
The anti-$k_T$ algorithm\,\cite{antiKt:2008gp} with a resolution parameter $R=0.6$ is used to reconstruct the jets. 
The jet reconstruction procedures used are similar to those of previous STAR analyses\,\cite{ STARCollaboration_2015_PrecisionMeasurement, Adam_2018_LongitudinalDoublespin, _2019_LongitudinalDoublespin, Abdallah_2021_LongitudinalDoublespin, STAR:2021mqa,Collins_2022}, except that the reconstructed $\Lambda$ and $\overline{\Lambda}$ candidates with invariant mass $1.08<m_{\Lambda(\overline{\Lambda})}<1.16$ $\GeV/c^2$ are included in the input particle list for jet reconstruction in addition to the TPC primary tracks and energy deposits in the BEMC and EEMC. 
To avoid double counting, the daughter tracks of $\Lambda$ or $\overline{\Lambda}$ candidates are removed from the input list. 
The energy deposits in $3\times3$ tower patches in the BEMC and EEMC with the central tower matched to a $\overline{p}$ daughter are also removed to correct the additional energy deposit due to annihilation of $\overline{p}$ with the EMC materials. 
The other jet reconstruction criteria remain the same. 
The TPC tracks are required to have $p_{T} \geq 0.2$ $\GeV/c$ and follow a $\pt$-dependent distance of closest approach to the event vertex as in Refs.\, \cite{Adamczyk_2012_LongitudinalTransverse, _2019_LongitudinalDoublespin,Abdallah_2021_LongitudinalDoublespin}.
The BEMC and EEMC towers are required to have a transverse energy $E_{T} \geq 0.2$ GeV. 
If a TPC track points to a BEMC or EEMC tower, a correction is applied to the tower $E_T$ to avoid double counting\,\cite{Adam_2018_LongitudinalDoublespin,
_2019_LongitudinalDoublespin,Abdallah_2021_LongitudinalDoublespin}.
The neutral energy fraction in the jet is required to be smaller than 0.95 \cite{_2019_LongitudinalDoublespin,Abdallah_2021_LongitudinalDoublespin,STAR:2021mqa,Collins_2022}.
To be included in further analysis, jets are required to have pseudorapidity relative to the event vertex in the range $-1.0<\eta_{jet}<1.0$ and relative to the center of STAR in the range $-0.7<\eta_{det}<0.9$. 
The reason for asymmetric $\eta_{det}$ is due to the EEMC acceptance, which only covers one side of STAR. 
Finally, the reconstructed jets are corrected for underlying-event contributions using the off-axis cone method\,\cite{offaxis_cone}. Jets with $p_T^{jet}>$ 5 GeV/$c$ after the correction are kept for further analysis.

The invariant mass distributions of the $\Lambda$ and $\overline{\Lambda}$ candidates after the above selection cuts with $1< \pth < 8$ GeV$/c$ and $|\etah|<1.2$ are shown in Fig.\,\ref{fig:im_L}.  
The bin counts under the signal mass windows are used to obtain the raw yields of $\Lambda$ and $\overline{\Lambda}$  candidates.
 The signal mass windows have been chosen to be about twice that of the fitted mass peak width. 
Approximately $1.56\times10^6$ $\Lambda$ and $1.67\times10^6$ $\overline{\Lambda}$ candidates in the longitudinal spin configuration, and $1.81\times10^6$ $\Lambda$ and $1.95\times10^6$ $\overline{\Lambda}$ candidates in the transverse spin configuration, are kept as the signal for further analysis.
The larger yield of $\overline{\Lambda}$ than $\Lambda$ is due to a bias in the jet patch trigger resulting from the additional energy deposit in the calorimeters associated with the annihilation of the antiproton daughter from $\overline{\Lambda}$ decay.
The slightly larger hyperon yield in the transverse spin configuration, compared to the longitudinal one, is related to different prescale factors for JP1 triggers in the two data sets, although their integrated luminosities are almost the same.

The residual background fraction under the mass peak is estimated by the side-band method\,\cite{Adam:2018kzl}, which sums the side-band regions on the left and right sides of the mass peak and then normalizes to the width of the signal window.
The estimated background fraction ranges from 6\% to 10\% among different bins. The mass window ranges of signal and side-band in each hyperon $\pth$ bin for spin transfer coefficient measurements as a function of $\pth$ are summarized in Table\,\ref{table:mass_window_pt}.

The spin transfer measurements reported here are for all detected $\Lambda$ and $\bar{\Lambda}$.  The embedded simulations described below predict that approximately 50\% of the $\Lambda$ and $\bar{\Lambda}$ are directly produced, while the remaining 50\% are decay products of $\Sigma^0$, $\Xi$, and other heavier baryons. 
Several of the theoretical models do take into account the decay contributions\,\cite{deFlorian:1998ba,Xu:2002hz,Xu:2005ru,Chen:2007tm}.

\begin{table*}[hbtp]

\centering
\caption{Summary of $\pth$-dependent mass windows for the hyperon signal region and side-bands.}
\centering
\footnotesize
\setlength{\tabcolsep}{5.5mm}{
\resizebox{0.85\linewidth}{!}{
\begin{tabular}{ c | c | c | c } \hline
    \multicolumn{4}{c}{Side-band and signal mass windows region ($\GeV/c^2$)}\\ \hline
        $\pth(\GeV/c)$ & left side-band  & signal window  & right side-band  \\ \hline
    1.0 $-$ 2.0 & (1.091, 1.106) & (1.111, 1.119)  & (1.124, 1.139)  \\ \hline
    2.0 $-$ 3.0 & (1.090, 1.105) & (1.110, 1.121)  & (1.126, 1.141) \\ \hline
    3.0 $-$ 4.0 & (1.087, 1.102) & (1.109, 1.123)  & (1.130, 1.145) \\ \hline
    4.0 $-$ 5.0 & (1.085, 1.100) & (1.108, 1.124)  & (1.132, 1.147) \\ \hline
    5.0 $-$ 6.0 & (1.084, 1.099) & (1.107, 1.126)  & (1.134, 1.149) \\ \hline
    6.0 $-$ 8.0 & (1.080, 1.095) & (1.105, 1.129)  & (1.139, 1.154) \\ \hline
\end{tabular}
}
}
\label{table:mass_window_pt}
\end{table*}

\begin{table*}[hbtp]

\centering
\caption{Summary of $z$-dependent mass windows for the hyperon signal region and side-bands.}
\centering
\footnotesize
\setlength{\tabcolsep}{6.5mm}{
\resizebox{0.85\linewidth}{!}{
\begin{tabular}{ c | c | c | c } \hline
    \multicolumn{4}{c}{Side-band and signal mass windows region ($\GeV/c^2$)}\\ \hline
        $z$ & left side-band  & signal window  & right side-band\\ \hline
    0.0 $-$ 0.1 & (1.091, 1.106) & (1.111, 1.119)  & (1.124, 1.139) \\ \hline
    0.1 $-$ 0.2 & (1.091, 1.106) & (1.111, 1.119)  & (1.124, 1.139) \\ \hline
    0.2 $-$ 0.3 & (1.089, 1.104) & (1.111, 1.120)  & (1.127, 1.142) \\ \hline
    0.3 $-$ 0.5 & (1.087, 1.102) & (1.110, 1.122)  & (1.130, 1.145) \\ \hline
    0.5 $-$ 0.7 & (1.085, 1.100) & (1.108, 1.124)  & (1.132, 1.147) \\ \hline
    0.7 $-$ 1.0 & (1.082, 1.097) & (1.107, 1.126)  & (1.136, 1.151) \\ \hline
\end{tabular}
}
}
\label{table:mass_window_z}
\end{table*}

\subsection{Jet momentum fraction carried by hyperon}

As mentioned in the introduction, the polarized fragmentation function can be better constrained by measuring the spin transfer  coefficient as a function of the jet momentum fraction $z$ carried by the $\Lambda$ or $\overline{\Lambda}$, which is defined as, 
\begin{equation}
  \centering
  z
  \equiv
   \frac{\overrightarrow{p}_{\Lambda}\cdot \overrightarrow{p}_{jet}}{|\overrightarrow{p}_{jet}|^2},
  \label{eq:z_def}
\end{equation}
 where $\overrightarrow{p}_{\Lambda}$ and $\overrightarrow{p}_{jet}$ are the momenta of the hyperon and jet, respectively. 
 As described in the previous subsection, hyperons are reconstructed from TPC tracks with good momentum precision (1-2 percent). The jets are reconstructed from TPC tracks, EMC energy deposits, and $\Lambda$ or $\overline{\Lambda}$ candidates, and the obtained jets at this level (before any correction for detector effects) are referred to as ``detector jets".  However, the true $z$ in Eq.\,(\ref{eq:z_def}) should be obtained with the jet momentum reconstructed with all the produced particles during the hadronization of a parton, which is referred to as a ``particle jet". 
 Correspondingly, the momentum fraction $z$ calculated using the jet momentum at the detector level or particle level with Eq.(\,\ref{eq:z_def}) is referred to as ``detector $z$" or ``particle $z$". The minimum and maximum hyperon $\pt$ cuts are removed for the spin transfer coefficient measurements as a function of the momentum fraction $z$. The signal mass window and the side-band regions in each detector $z$ bin are summarized in Table\,\ref{table:mass_window_z}.

\begin{figure}[htbp]
\centering
\includegraphics[width=1.0\linewidth]{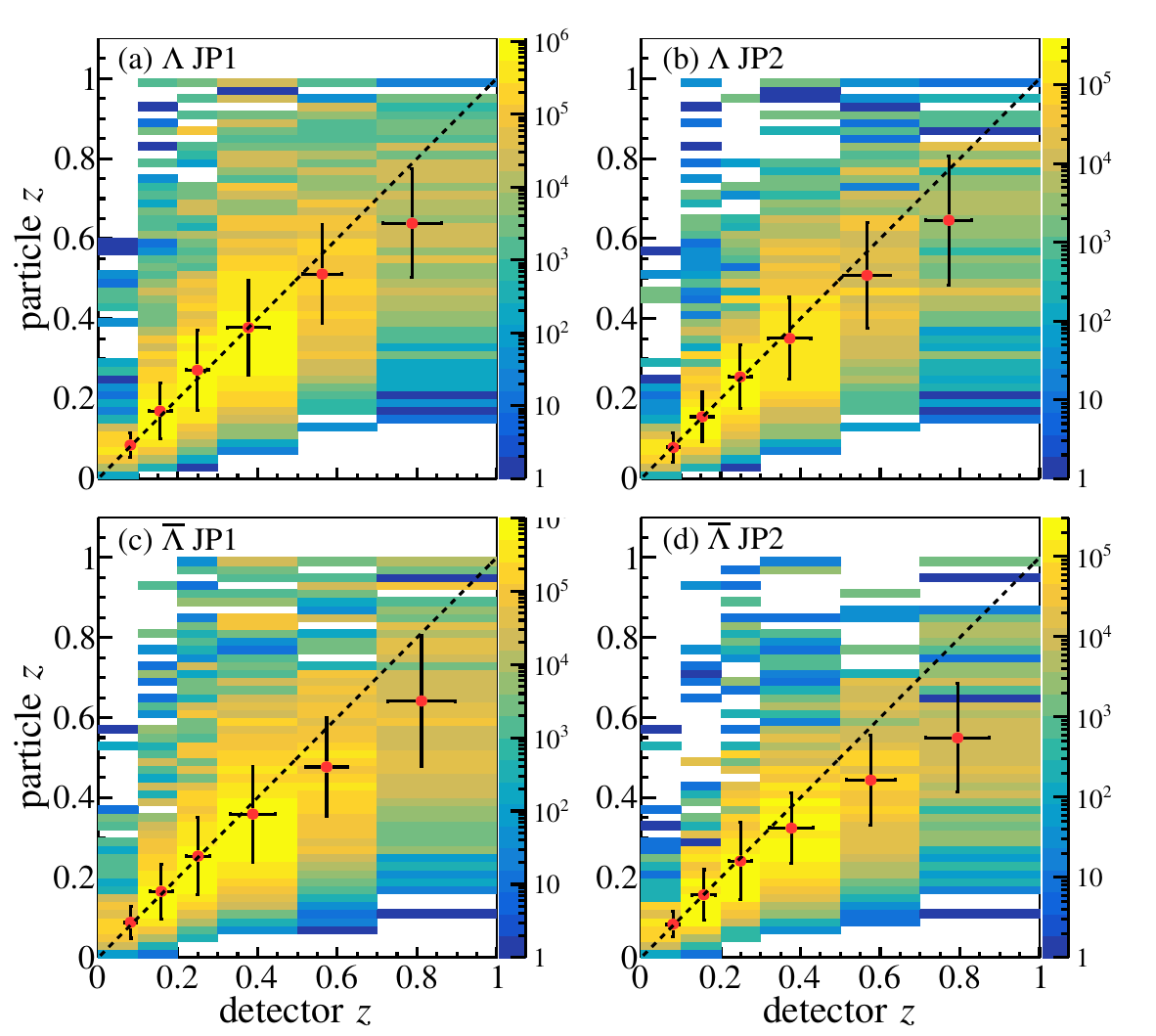}
\caption{The correlation of jet momentum fraction $z$ carried by $\Lambda$ (upper panels) and $\overline{\Lambda}$ (lower panels) at particle level and detector level, for jet triggers JP1 (left) and JP2 (right). The red points give the mean values of ``detector $z$" and ``particle $z$" in each bin while the error bars represent the standard derivations. The dashed lines at $y=x$ are for guidance.   }
\label{fig:z_corr}
\end{figure}

In order to compare the experimental results with theoretical predictions, which are calculated at the particle level, a correction needs to be applied to the detector $z$ in our measurement. 
The correction has been obtained from Monte-Carlo (MC) events that are generated with PYTHIA6\,\cite{pythia6}, then passed through the full simulation of the STAR detector based on the GEANT3\,\cite{geant3} framework and embedded into zero-bias events collected at STAR to account for the background environment of real data.  
The same reconstruction procedures and same cuts used for data are applied to the MC events for both hyperon selection and jet reconstruction.
To associate the jets and hyperons at the particle level to the detector level, a cut on their separation in $\eta$ and $\phi$ space is applied: $\Delta R < 0.5$ for the jet and $\Delta R < 0.05$ for the hyperon, with $\Delta R\equiv\sqrt {(\Delta \phi)^2 + (\Delta \eta)^2}$.
Figure \ref{fig:z_corr} shows the correlation of particle $z$ and detector $z$ for $\Lambda$ and $\overline{\Lambda}$ from the embedded simulation of $pp$ collisions with JP1 and JP2 triggers at $\sqrt s$=200 GeV.
The average values in each detector $z$ bin are also shown.
No clear difference is seen for JP1 and JP2 triggers within uncertainties.
The correction, $\delta z$, is calculated as the difference of particle $z$ and detector $z$ in each detector $z$ bin.
Then the corresponding $\delta z$ for each detector $z$ bin is applied to each data point, and thus the momentum fraction value at detector level is corrected to particle level.
It is seen that the $\delta z$ correction of $\overline\Lambda$ is slightly larger than that of $\Lambda$. This is related to the trigger bias due to antiproton annihilation within the EM calorimeters under the same jet transverse energy threshold.

\subsection{Spin transfer coefficient extraction}
\subsubsection{Extraction of $\DLL$}
As in the previous measurement\,\cite{Abelev:2009xg}, the longitudinal spin transfer coefficient $D_{LL}$ is extracted from the asymmetry of hyperon yields in a small $\cos\theta^*$ interval when the proton beam is positively and negatively polarized:
\begin{equation}
D_{LL}=\frac{1}{\alpha_{\Lambda(\overline{\Lambda})} P_\mathrm{beam} \left<\cos \theta^*\right>} \frac{N^+ -\mathcal{R} N^-} {N^+ + \mathcal{R} N^-},
\label{eq_dll}
\end{equation}
where $N^+$ $(N^-)$ is the number of $\Lambda$ or $\overline{\Lambda}$ candidates in the $\cos\theta^*$ interval when the beam helicity is positive (negative), and $\alpha_{\Lambda}=0.732\pm0.014$\,\cite{PDG}, $\alpha_{\overline{\Lambda}} = -\alpha_{\Lambda}$ (assuming no CP violation). $P_{\mathrm{beam}}$ is the beam polarization and $\left<\cos\theta^*\right>$ is the average value of $\cos\theta^*$ in the interval. 
$\mathcal{R}$ denotes the luminosity ratio for the two beam polarization states.
At RHIC, both beams are polarized, and the single spin yields $N^+$ and $N^-$ are obtained by summing over the opposing-beam spin, weighted by the corresponding relative luminosities\,\cite{Adam:2018kzl}.
The relative luminosities are measured with the VPD\,\cite{VPD} and the ZDC\,\cite{ZDC}.
In Eq.\,(\ref{eq_dll}), the acceptance cancels as it remains the same when flipping the beam polarization\,\cite{Abelev:2009xg} in a small $\cos\theta^*$ interval.
The raw spin-transfer values $\DLL^{raw}$ are first obtained with Eq.\,(\ref{eq_dll}) using the number of hyperon counts under the mass peak, then averaged over the entire $\cos\theta^*$ range.
Figure \ref{fig:costheta}(a) shows an example of $\DLL^{raw}$ extraction as a function of
 $\cos\theta^*$ with $3< \pth <4$ $\GeV/c$ and $0<\etah <1.2$. 
 
A correction is applied to subtract the contribution from the residual background (similar corrections are also applied to the statistical uncertainty):

\begin{align}
    &\DLL =\frac{\DLL^{raw}-r\DLL^{bg}}{1-r},\label{eq_corr_bkg} \\
    &\delta \DLL = \frac{\sqrt{(\delta\DLL^{raw})^2+(r\delta\DLL^{bg})^2}}{1-r},\label{eq_corr_bkg_err}
\end{align}
where $\DLL^{bg}$ is the spin transfer value obtained from the side-band region, and $r$ is the residual background fraction under the mass peak calculated using the side-band method\,\cite{Adam:2018kzl}.
$\DLL^{bg}$ is found to be consistent with zero within uncertainties.
The spin transfer results from each of the two beams were found to be consistent with each other, and their weighted average was used for the final result.
%

\subsubsection{Extraction of $\DTT$}

\begin{figure}[htbp]
    \includegraphics[width=1\linewidth]{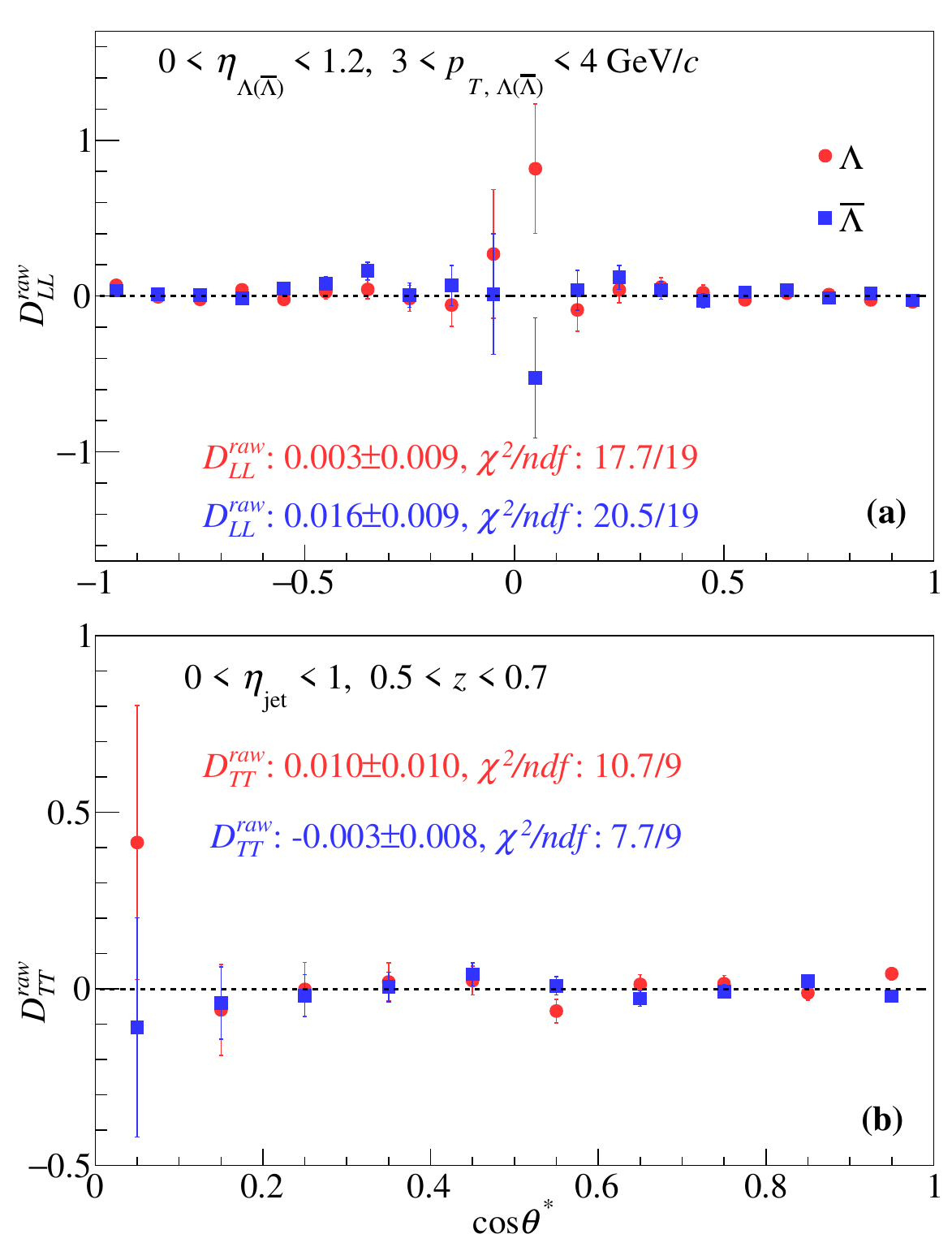}
    \caption{(a) Longitudinal spin transfer coefficient $\DLL^{raw}$ of $\Lambda$ and $\overline{\Lambda}$ as a function of $\mathrm{cos}\,\theta^{*}$ for hyperons with $3 < \pth < 4$ $\GeV/c$. (b) Transverse spin transfer coefficient $\DTT^{raw}$ of $\Lambda$ and $\overline{\Lambda}$ as a function of $\mathrm{cos}\,\theta^*$ for hyperons with momentum fraction $0.5 < z < 0.7$.}
    \label{fig:costheta}
\end{figure}

To minimize the systematic effects associated with detector acceptance and luminosity ratios, the transverse spin transfer coefficient $\DTT$ is extracted 
using the same cross-ratio method as the previous publication\,\cite{Adam:2018dtt}:

\begin{widetext}
\begin{equation}
  \DTT = 
  \frac{1}{\,\alpha_{\Lambda(\overline{\Lambda})} P_{\mathrm{beam}} \left<\cos{\theta^{*}}\right>\,}
  \frac
    {\,\sqrt{N^{\uparrow}(\cos{\theta^{*}})N^{\downarrow}(-\cos{\theta^{*}})\,}
  -\sqrt{N^{\downarrow}(\cos{\theta^{*}})N^{\uparrow}(-\cos{\theta^{*}})\,}\, }
  {\,\sqrt{N^{\uparrow}(\cos{\theta^{*}})N^{\downarrow}(-\cos{\theta^{*}})\,}
  +\sqrt{N^{\downarrow}(\cos{\theta^{*}})N^{\uparrow}(-\cos{\theta^{*}})\,}\, },
  \label{eq:dttAsy}
\end{equation}
\end{widetext}
where $N^{\uparrow}$ ($N^{\downarrow}$) is the $\Lambda$ or $\overline{\Lambda}$ yield in the corresponding $\cos\theta^*$ bin when the proton beam is 
polarized upward (downward).
The acceptance and the luminosity ratio between $N^{\uparrow}$ and $N^{\downarrow}$ cancel in this cross-ratio asymmetry. 
As mentioned in the introduction, the transverse polarization direction of the outgoing quark is used to obtain $\theta^{*}$\,\cite{Adam:2018dtt}. In practice, the reconstructed jet axis is taken as the direction of the outgoing quark (see Fig. 1) in applying the rotation between the transverse polarization directions of the incoming and outgoing quarks along the normal direction of the partonic scattering plane\,\cite{Collins:1993kq,Adam:2018dtt}.
Figure\,\ref{fig:costheta}(b) shows an example of $\DTT^{raw}$ as a function of $\mathrm{cos}\,\theta^*$ for $\Lambda$ and $\overline{\Lambda}$ with $0.5 < z < 0.7$ and $0<\eta_{jet}<1.0$.
The final $\DTT$ results are corrected for residual background using equations similar to Eqs.\,(\ref{eq_corr_bkg}) and (\ref{eq_corr_bkg_err}).

\subsection{Systematic uncertainties}
The following sources of systematic uncertainties are considered and discussed in more detail below. 
Uncertainties in the $\Lambda$ decay parameter and the beam polarization are fully correlated for all the $D_{LL}$ and $D_{TT}$ results in different kinematic bins. The uncertainties in the luminosity ratio only contribute to $D_{LL}$ measurements and are also fully correlated. 
Additional uncertainties in determining the residual background fraction and
introduced by the trigger conditions fluctuate point-to-point in hyperon $\pt$ and $z$. 
\begin{itemize}[leftmargin=0pt, itemindent=1em]
    \item \text{Hyperon decay parameter}: 
    The decay parameter of $\Lambda$, $\alpha_{\Lambda}=0.732\pm0.014$\,\cite{PDG} with $\alpha_{\overline{\Lambda}}=-\alpha_{\Lambda}$, has a relative uncertainty of about $1.9\%$, which is applied to the measured spin transfer coefficients as an overall scale uncertainty. 
    \item \text{Beam polarizations}:
The relative uncertainties of the beam polarizations during 2015 are about $3\%$ for both longitudinally and transversely polarized beam configurations\,\cite{PolBeam_RHIC}, which are also applied to $\DLL$ and $\DTT$ as a scale uncertainty.
    \item \text{Luminosity ratio}:
The uncertainty of the luminosity ratio $\mathcal{R}$ is found to be about 0.0007, and applied to the $\DLL$ measurements through Eq.\,(\ref{eq_dll}). Estimated as in Ref.\,\cite{Abdallah_2021_LongitudinalDoublespin}, the corresponding systematic uncertainty to $\DLL$ is about 0.0020. There is no such uncertainty for the $\DTT$ measurement as the luminosity ratio cancels in the cross-ratio method.
    \item \text{Residual background}:
    The uncertainty of the residual background fraction $r$ in Eq.\,(\ref{eq_corr_bkg}) is taken as another source of systematic uncertainty. In addition to the side-band method, the fitting method with a Gaussian$+$linear function was also used to estimate the background fraction, and the corresponding difference of the extracted spin transfer values was taken as the systematic uncertainty of $\DLL$ and $\DTT$. Overall, this part is quite small, up to 0.0010 (0.0007) for $\DLL$ ($\DTT$) at high $\pth$, which is less than 10\% of the statistical uncertainty.
%
    \item \text{Trigger bias}:
    The data sets used in this analysis were recorded with jet-patch trigger conditions, which may bias the spin transfer coefficient measurements by preferentially selecting certain processes leading to $\Lambda$ and $\overline{\Lambda}$ production as mentioned in previous publications\,\cite{Adam:2018kzl,Adam:2018dtt}. Similar to previous measurements, this potential bias is 
    studied with the MC simulation events generated with PYTHIA6\,\cite{pythia6} and the STAR detector response package based on GEANT3\,\cite{geant3}. The biases introduced by the trigger conditions are evaluated from the difference of $\DLL$ and $\DTT$ results with a model\,\cite{Xu:2005ru} before and after applying the trigger conditions in the MC simulation. The trigger bias is the dominant source of systematic uncertainties for both the $\DLL$ and $\DTT$ measurements. It increases with $\pth$ and $z$ in general, and is as large as 0.0131 (0.0088) for $\DLL$ ($\DTT$).
\end{itemize}

\section {Results and discussion}
\setlength{\parskip}{0cm}
\subsection{Results for $\DLL$}
\subsubsection{$\DLL$ results as a function of the hyperon $\pth$}
\begin{figure}[htbp]
\includegraphics[width=1\linewidth]{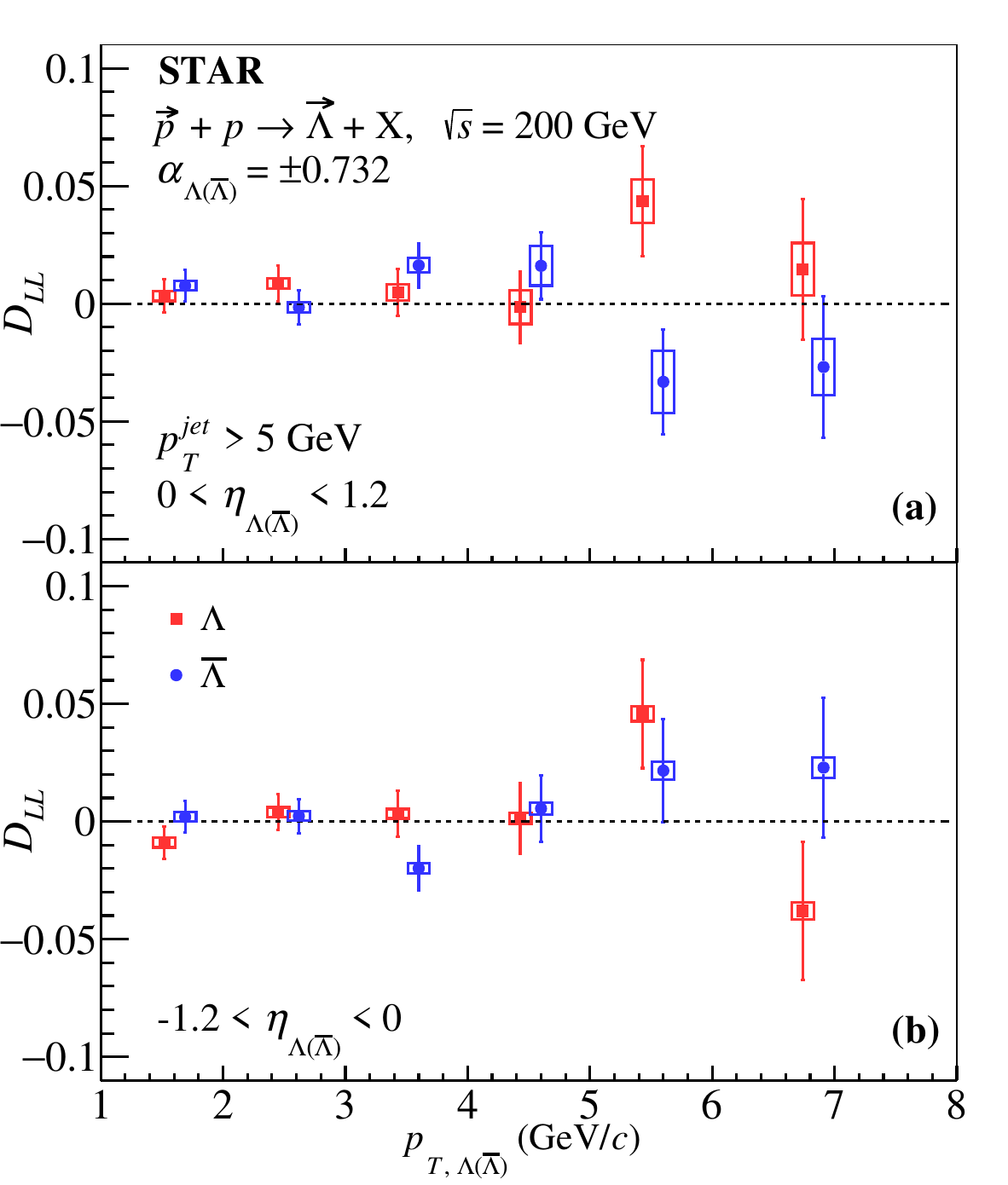}
\caption{Longitudinal spin transfer coefficient $\DLL$ of $\Lambda$ and $\overline{\Lambda}$ as a function of hyperon $p_T$ in proton-proton collisions at $\sqrt{s}=200$ $\GeV$. The top and bottom panels show the results for positive and negative hyperon $\eta$ regions, respectively. The vertical bars and boxes indicate the statistical and systematic uncertainties, respectively. The $\overline{\Lambda}$ results have been slightly offset horizontally for clarity.}
\label{fig:DLL_vs_pt}
\end{figure}

\setlength{\parskip}{0pt}
The longitudinal spin transfer coefficient, $\DLL$, as a function of hyperon $\pth$ in proton-proton collisions at $\sqrt s=200$ $\GeV$ is shown in Fig.\,\ref{fig:DLL_vs_pt}. 
The top panel shows the results with positive hyperon $\eta$ of $0<\etah<1.2$ and the bottom panel with $-1.2<\etah<0$, with positive pseudorapidity defined along the momentum direction of the polarized beam.
The spin transfer in the backward region (negative $\etah$) is expected to be significantly smaller than that in the forward region (positive $\etah$) relative to the polarized proton beam\,\cite{deFlorian:1998ba,Boros:2000ya,Ma:2001na,Xu:2002hz,Xu:2005ru}.
The vertical bars represent the statistical uncertainties, and the systematic uncertainties are shown in boxes. 
The results show no evidence for a difference between $\Lambda$ and $\overline{\Lambda}$ within uncertainties.

\begin{figure}[h!]
\includegraphics[width=1\linewidth]{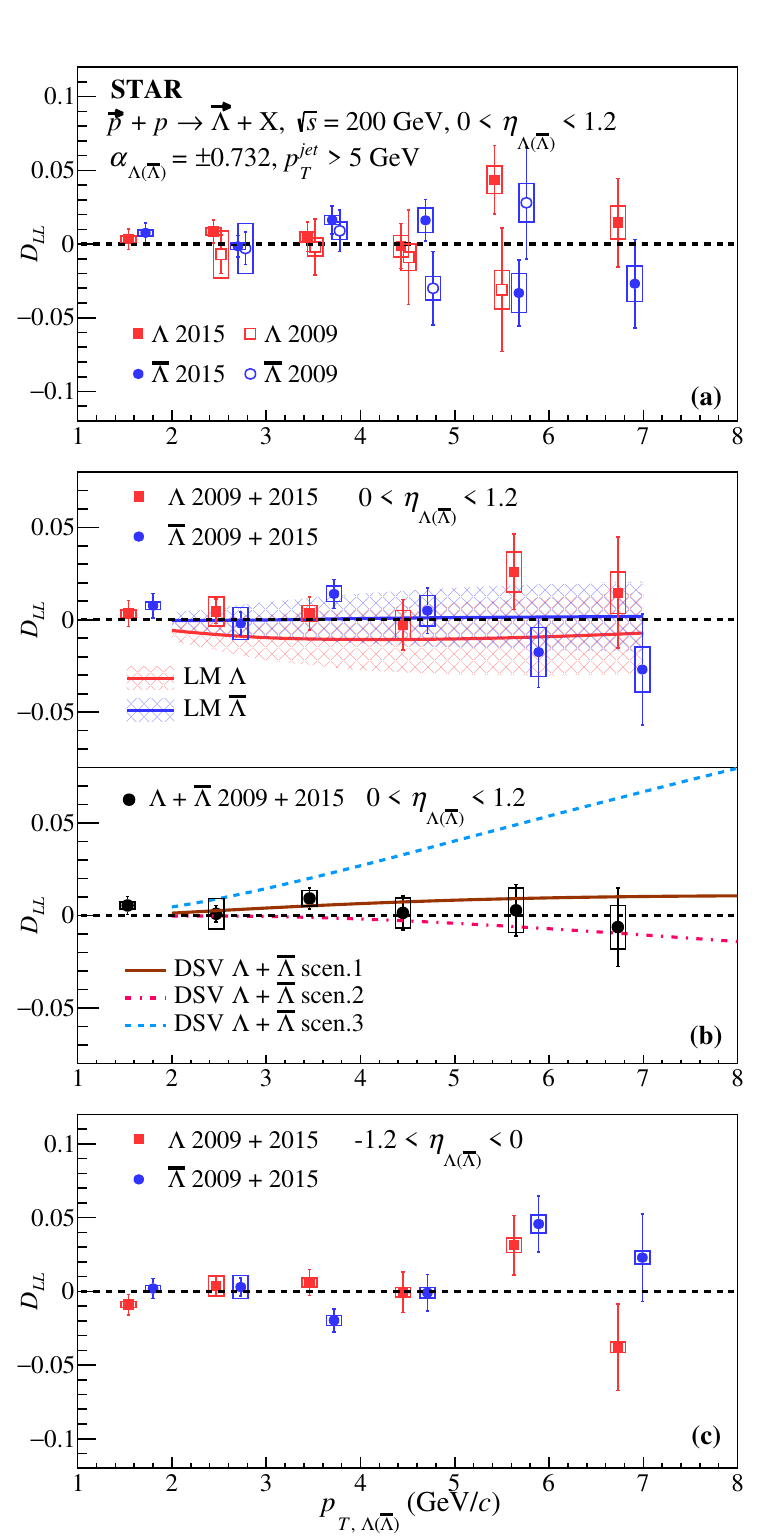}
\caption{(a) Comparison of longitudinal spin transfer coefficient $D_{LL}$ as a function of the hyperon $\pth$ for positive $\eta$ with previously published results \cite{Adam:2018kzl}. (b) Upper sub-panel: combined results of $D_{LL}$ for positive $\eta$ from current and previous measurements, in comparison with theoretical prediction\,\cite{Liu:2019xcf}; Lower sub-panel: the $\Lambda + \overline{\Lambda}$ combined results, in comparison with theoretical predictions\,\cite{deFlorian:1998ba,DLL_pt_curve}.
(c) Combined results of $D_{LL}$ for negative $\eta$ from current and previous measurements. The previously published results in panel (a) and the results of $\overline{\Lambda}$ in all panels are slightly shifted for clarity. 
}
\label{fig:DLL_run15_vs_run9}
\end{figure}

Figure\,\ref{fig:DLL_run15_vs_run9}(a) shows the comparison of $\DLL$ results obtained here in the positive $\eta$ range with previously published results based on STAR data taken in 2009\,\cite{Adam:2018kzl}. 
We note that the previous results are rescaled with $\alpha_{\Lambda}=0.732\pm0.014$ here.
The current $D_{LL}$ results are consistent with the results previously published by STAR, and the statistics in this measurement are about 2 times larger than those in the previous publication. Similar agreement is found for the measurements at negative $\etah$.
We calculate the statistical average of the new measurements and the previous ones, with systematic uncertainties taken as their weighted average based on the hyperon yields in different years.
The combined results of $D_{LL}$ from these two measurements are shown in Fig.\,\ref{fig:DLL_run15_vs_run9}(b) for positive $\eta$ and in Fig.\,\ref{fig:DLL_run15_vs_run9}(c) for negative $\eta$.

Theoretical predictions ``LM" from Ref.\,\cite{Liu:2019xcf}, which considers $\DLL$ with $\Lambda$ and $\overline{\Lambda}$ separately and uses STAR 2009 results as input, are in general consistent with the combined $D_{LL}$ results in upper sub-panel of Fig.\,\ref{fig:DLL_run15_vs_run9}(b). 
Predictions ``DSV" from Refs.\,\cite{deFlorian:1998ba,DLL_pt_curve}, which calculates $\DLL$ with $\Lambda$ and $\overline{\Lambda}$ combined, are compared with the $\Lambda + \overline{\Lambda}$ combined results in lower sub-panel of Fig.\,\ref{fig:DLL_run15_vs_run9}(b). 
Here different scenarios of ``DSV" curves are related to different assumptions for the polarized fragmentation functions\,\cite{deFlorian:1998ba}, which are still poorly constrained by experimental data. 
``DSV $\Lambda + \overline{\Lambda}$ scen.1” is based on the expectations from the naive quark model, where only strange quarks can contribute to the $\Lambda$ polarization during the fragmentation processes, while in ``DSV $\Lambda + \overline{\Lambda}$ scen.2” a sizable negative contribution from $u$ and $d$ quarks to $\Lambda$ polarization is assumed, similar to the DIS picture of nucleon spin\,\cite{deFlorian:1998ba}.
The ``DSV $\Lambda + \overline{\Lambda}$ scen.3'' is based on an extreme assumption that the polarized fragmentation functions are independent of quark flavor, i.e., $u$, $d$ and $s$ quarks contribute equally\,\cite{deFlorian:1998ba}. 
The STAR results are consistent with ``DSV $\Lambda + \overline{\Lambda}$ scen.1” and ``DSV $\Lambda + \overline{\Lambda}$ scen.2” predictions within uncertainties.
The data points lie below the ``DSV $\Lambda + \overline{\Lambda}$ scen.3” predictions, and the $\chi^2/ndf$ of combined $\Lambda + \overline{\Lambda}$ $\DLL$ results with this scenario is 24.2/5. The large $\chi^2$ value indicates that this extreme assumption is strongly disfavored. 


\begin{figure}[h]
\includegraphics[width=1\linewidth]{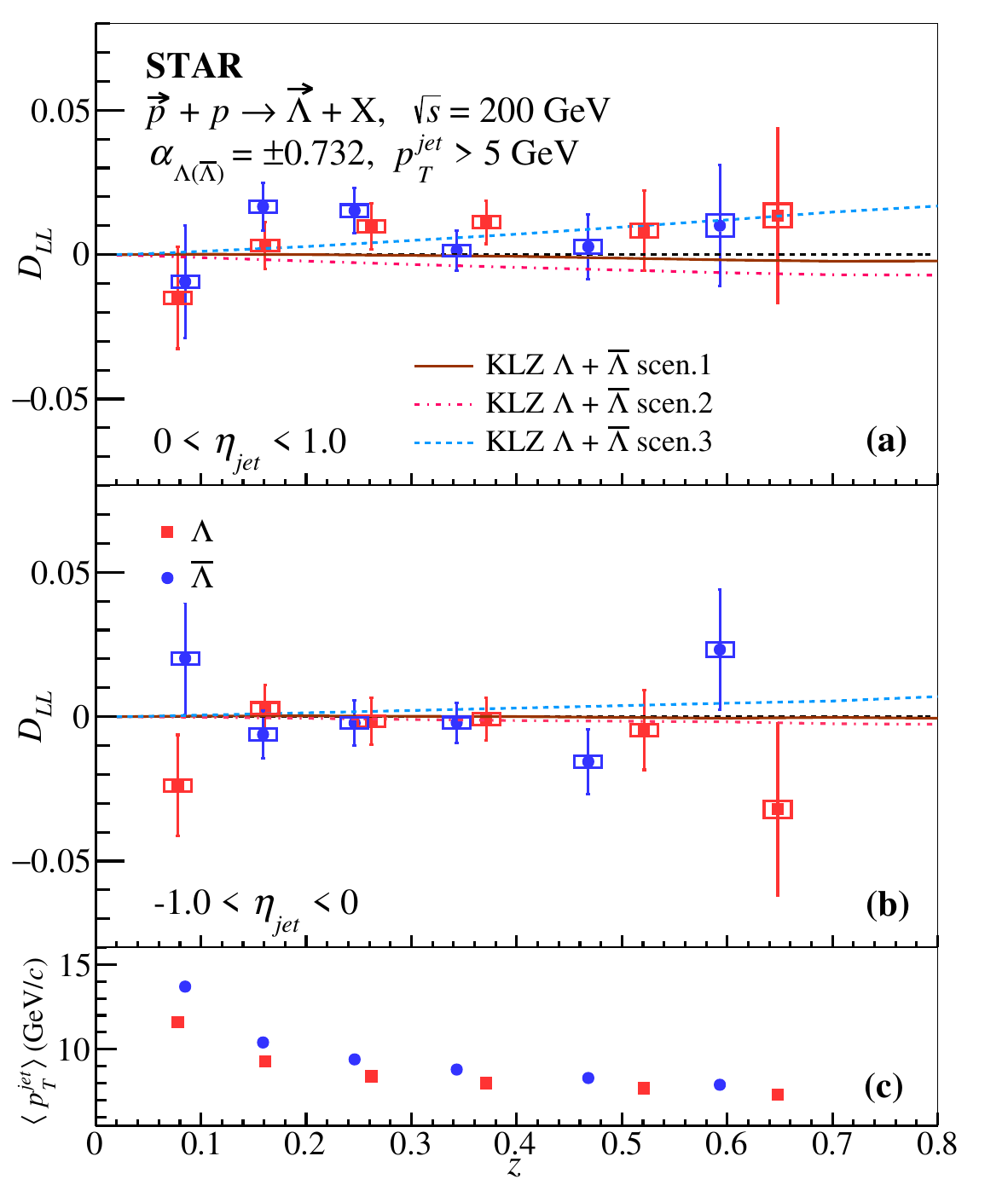}
\caption{Longitudinal spin transfer coefficient $\DLL$ as a function of the momentum fraction $z$ of the hyperon within a jet in proton-proton collisions at $\sqrt{s}=200$ $\GeV$ compared with theoretical calculations\,\cite{Kang:2020xyq}. Panels (a) and (b) show the results for positive and negative $\etaj$, respectively. The average jet $\pt$ at the particle level in each $z$ bin is shown in panel (c). 
Here the differences of $z$ value for $\Lambda$ and $\overline\Lambda$ along the horizontal axis reflect their average $z$ in that bin after the correction to particle level, not an artificial offset.
}
\label{fig:DLL_vs_z}
\end{figure}

\subsubsection{$D_{LL}$ results as a function of the momentum fraction $z$ in jets}
The longitudinal spin transfer coefficient $\DLL$ as a function of the momentum fraction $z$ in jets in proton-proton collisions at $\sqrt s =200$ GeV is shown in Fig.\,\ref{fig:DLL_vs_z}. 
The panels (a) and (b) show the results for positive and negative jet pseudorapidity $\etaj$.
Panel (c) shows the average jet $\pt$ at the particle level in each $z$ bin.
Here the differences of $z$ value for $\Lambda$ and $\overline\Lambda$ along the horizontal axis reflect their average $z$ in that bin after the correction to particle level.
This is the first measurement of the spin transfer coefficient $\DLL$ as a function of jet momentum fraction within a jet, and it provides a direct probe of the polarized fragmentation function of the $\Lambda$ hyperon.
The STAR results are compared with theoretical predictions ``KLZ" from Ref.\,\cite{Kang:2020xyq} as shown in Fig.\,\ref{fig:DLL_vs_z}. 
Three scenarios for the polarized fragmentation functions\,\cite{deFlorian:1998prd} are also used in these predictions.
As can be seen, the STAR results are consistent with the model calculations within uncertainties. 
The results for $\Lambda$ and $\overline{\Lambda}$ are also consistent with each other.
More statistics are needed, in particular for the high-$z$ region, to distinguish between the different scenarios.

\begin{figure}[h]
\includegraphics[width=1\linewidth]{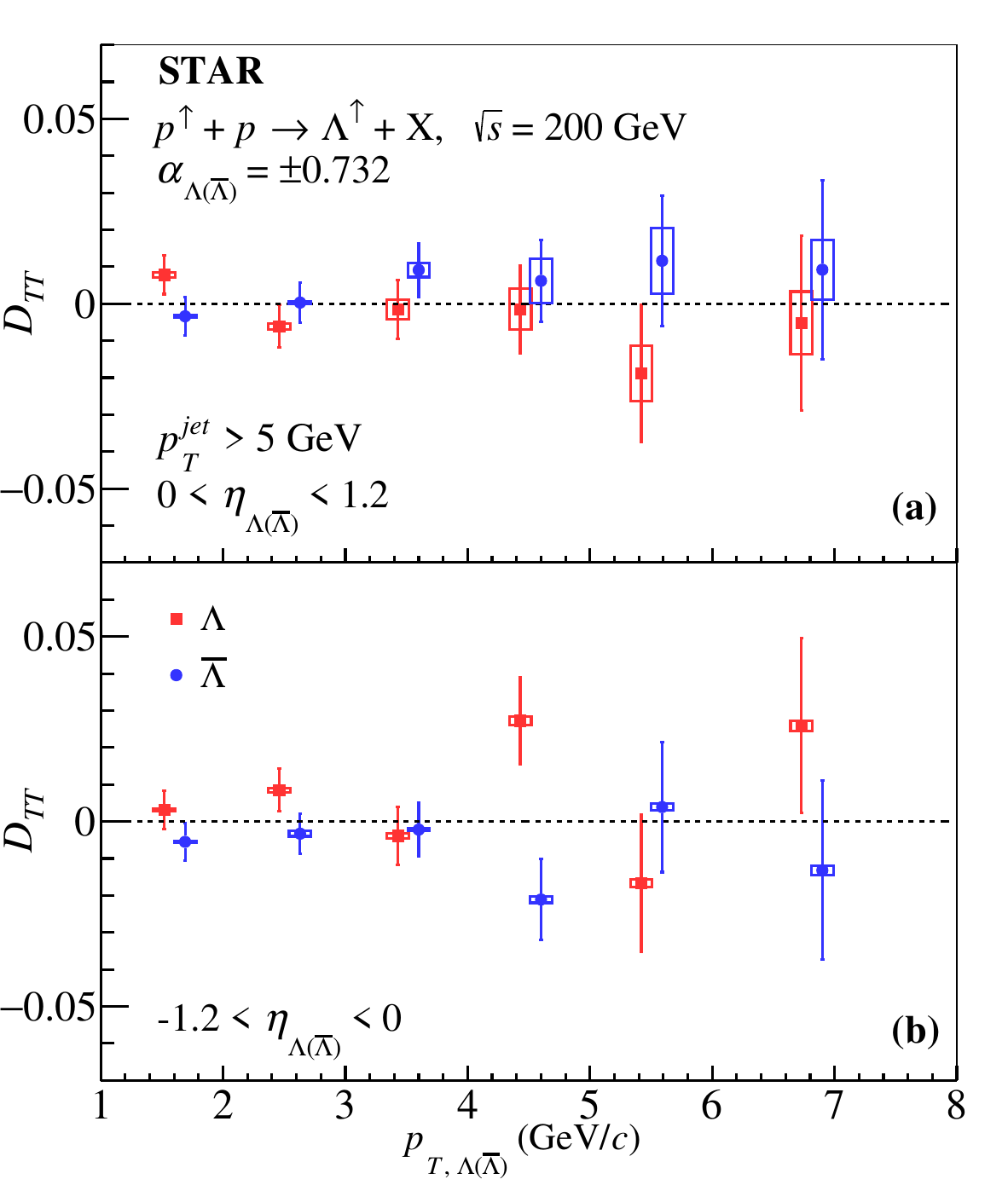}
\caption{Transverse spin transfer coefficient $\DTT$ as a function of hyperon $p_T$ in proton-proton collisions at $\sqrt{s}=200$ $\GeV$ at STAR. The top and bottom panels show the results for positive and negative $\etah$, respectively. The $\overline{\Lambda}$ results have been slightly offset horizontally for clarity.}
\label{fig:DTT_vs_pt}
\end{figure}
\subsection{Results for $\DTT$}

\subsubsection{$\DTT$ results as a function of hyperon $\pth$}

The transverse spin transfer coefficient $\DTT$ as a function of hyperon $p_T$ in proton-proton collisions at $\sqrt s=200$ $\GeV$ is shown in Fig.\,\ref{fig:DTT_vs_pt}.
Results are shown in two hyperon $\eta$ regions: $0<\etah<1.2$ (top panel) and $-1.2<\etah<0$ (bottom panel) with pseudorapidity defined with respect to the polarized beam. 
The $\DTT$ results for $\Lambda$ and $\overline{\Lambda}$ are consistent with each other within uncertainties.
\vspace{6pt}


\begin{figure}[htbp]
\includegraphics[width=1\linewidth]{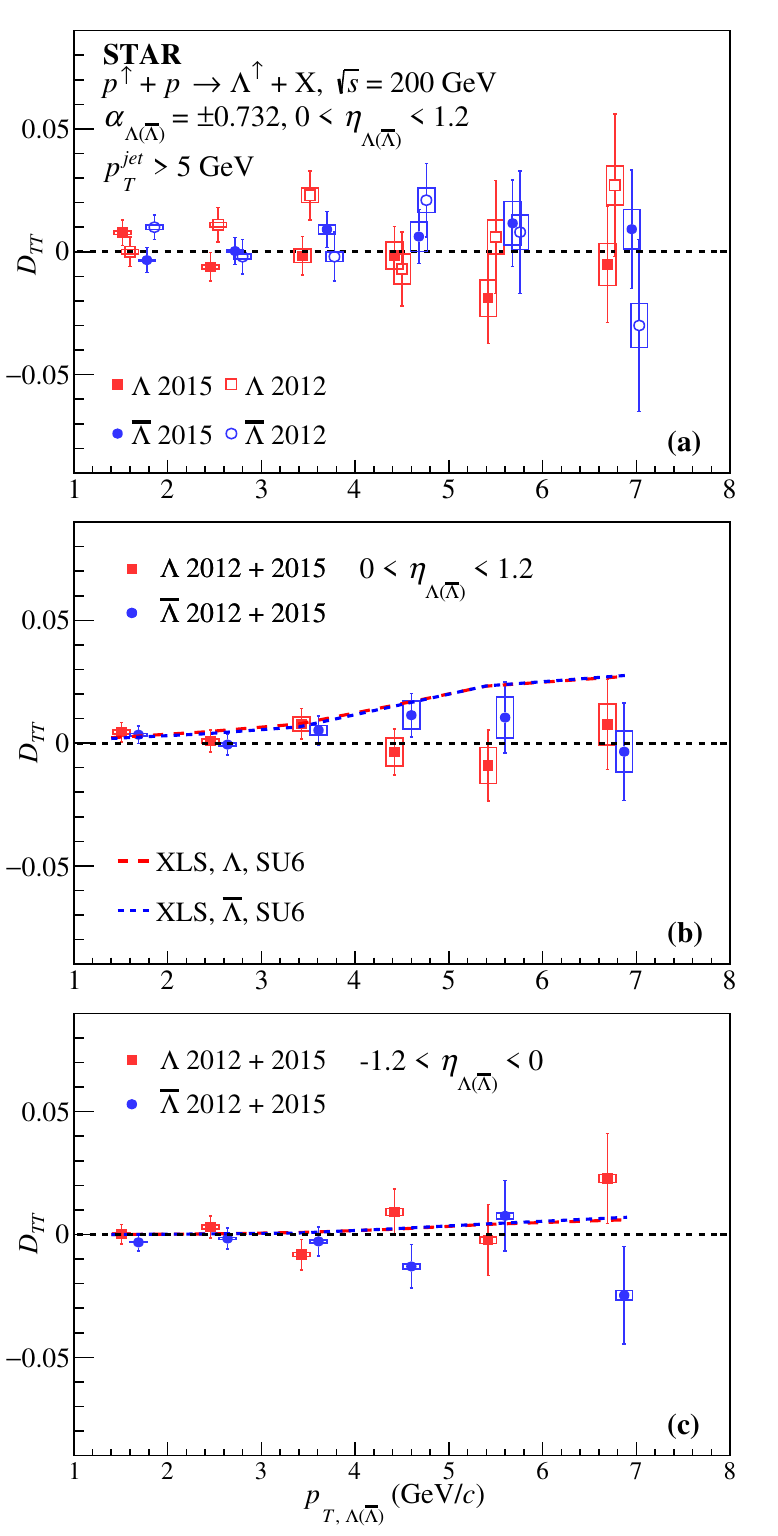}
\caption{(a) Comparison of transverse spin transfer coefficient $D_{TT}$ as a function of hyperon $\pth$ for positive $\eta$ with previously published results \cite{Adam:2018dtt}. (b) Combined results of $D_{TT}$ for positive $\eta$ from current and previous measurements, in comparison with theoretical predictions\,\cite{Xu:2005ru}. (c) Combined results of $D_{TT}$ for negative $\eta$ from current and previous measurements. The previously published results and the results of $\overline{\Lambda}$ are slightly shifted horizontally for clarity.}
\label{fig:DTT_run15_vs_run12}
\end{figure}

Figure\,\ref{fig:DTT_run15_vs_run12}(a) shows the comparison of $\DTT$ results with previously published results\,\cite{Adam:2018dtt} for positive $\etah$ based on STAR data taken in 2012.
We note that the previous results are rescaled with $\alpha_{\Lambda}=0.732\pm0.014$ here.
The $D_{TT}$ results in this analysis are consistent with the previous results\,\cite{Adam:2018dtt}, and the new measurement has a factor of 2 improvement in statistics compared to the previous one. Similar agreement is found for the measurements at negative $\etah$.
The combined results of $D_{TT}$ from these two measurements are shown in Fig.\,\ref{fig:DTT_run15_vs_run12}(b) for positive $\etah$ and in Fig.\,\ref{fig:DTT_run15_vs_run12}(c) for negative $\etah$.
Theoretical predictions ``XLS" from Ref.\cite{Xu:2005ru} with a simple assumption that the strange quark transversity is equal to its helicity distribution are also compared with the combined results.
In this model, the spin transfer coefficient in the positive $\etah$ region is expected to be larger than that in negative $\etah$ region. 
From the comparison in Fig.\,\ref{fig:DTT_run15_vs_run12}(b), 
the $D_{TT}$ results of $\Lambda$ and $\overline{\Lambda}$ at positive $\eta$ generally fall below the model predictions.
However, the current statistics are still limited, especially at high $\pt$. 
Small $\DTT$ results might indicate small transversely polarized fragmentation functions and/or small transversity of the strange quark and anti-quark inside the proton.

\subsubsection{$\DTT$ results as a function of the momentum fraction $z$ in jets}
\begin{figure}[htbp]
\includegraphics[width=1\linewidth]{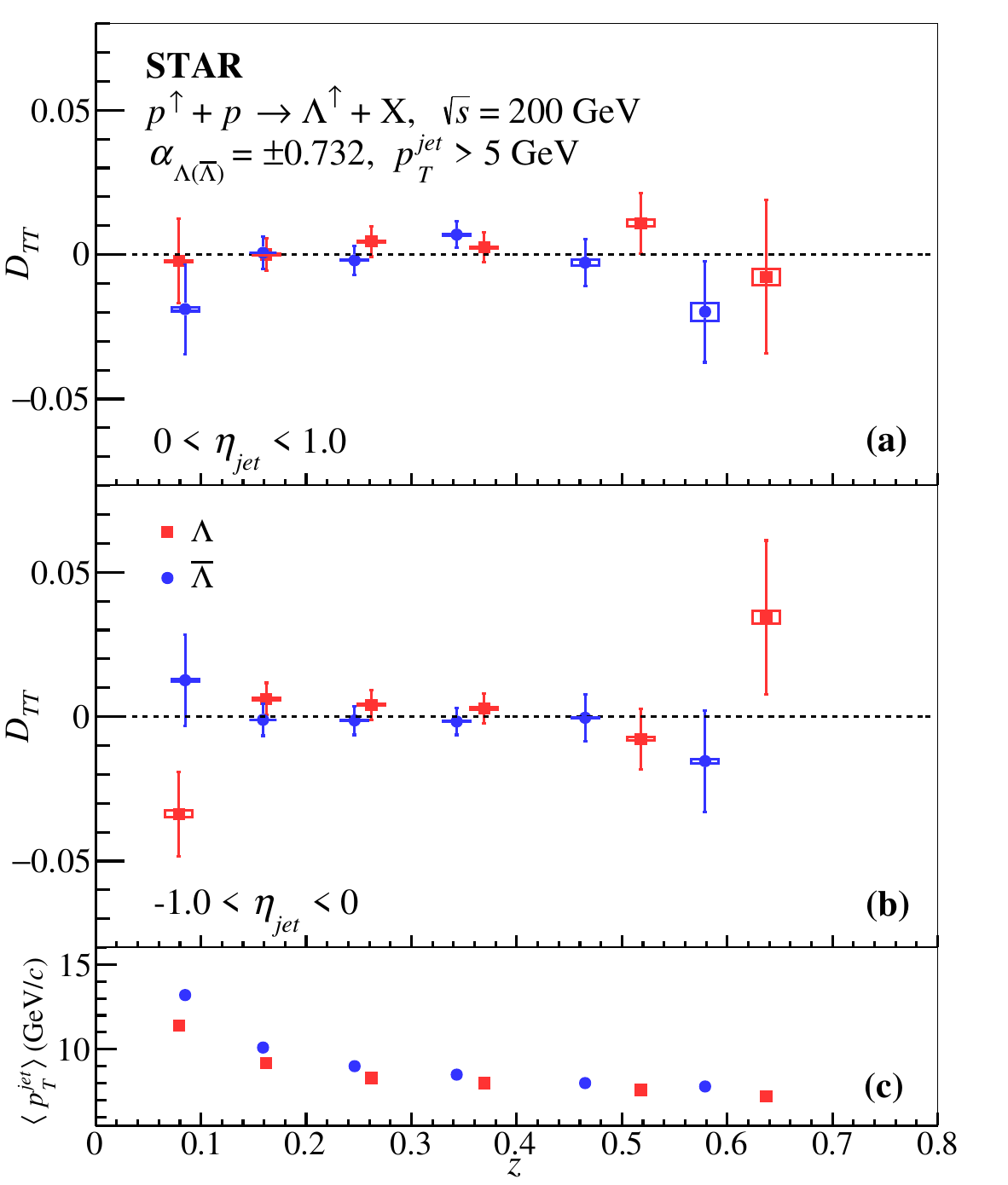}
\caption{Transverse spin transfer coefficient $\DTT$ as a function of the momentum fraction $z$ of the hyperon in a jet in proton-proton collisions at $\sqrt{s}=200$ $\GeV$. The panels (a) and (b) show the results for positive and negative $\etaj$, respectively.
The average jet $\pt$ at particle level in each $z$ bin is shown in panel (c). 
Here the differences of $z$ value for $\Lambda$ and $\overline\Lambda$ along the horizontal axis reflect their average $z$ in that bin after the correction to particle level, not an artificial offset.}
\label{fig:DTT_vs_z}
\end{figure}

Figure\,\ref{fig:DTT_vs_z} shows the first measurement of the transverse spin transfer coefficient $\DTT$ as a function of momentum fraction $z$ in jets in proton-proton collisions at $\sqrt s=200$ GeV. 
The top and middle panels show the results for positive and negative $\etaj$ ranges with respect to the polarized beam, while the bottom panel shows the average jet $\pt$ at particle level in the corresponding $z$ bin. 
Here the differences of $z$ value for $\Lambda$ and $\overline\Lambda$ along the horizontal axis reflect their average $z$ in that bin after the correction to particle level.
The results for $\Lambda$ and $\overline{\Lambda}$ are consistent with each other within uncertainties.
Currently there are no theoretical predictions for $\DTT$ as a function of $z$. 
These new $\DTT$ results as a function of $z$ will provide direct constraints on the transversely polarized fragmentation functions for $\Lambda$ and $\overline{\Lambda}$. More studies on hyperon transverse polarization are needed for a better understanding of both the transversity distribution and polarized fragmentation functions. 

\section {Conclusion}

Measurements of the spin transfer coefficients from a polarized proton beam to the produced $\Lambda$($\overline{\Lambda})$ hyperons in polarized proton-proton collisions can provide valuable information on proton spin structure related to the (anti-)strange quarks and the polarized fragmentation functions. 
The longitudinal spin transfer coefficient $\DLL$ to $\Lambda$ and $\overline{\Lambda}$ hyperons provides connections to the helicity distributions and the longitudinally polarized fragmentation functions, while the transverse spin transfer coefficient $\DTT$ is related to the transversity distribution and transversely polarized fragmentation functions.
\vspace{4pt}

\par 
In this paper, we report improved measurements of both $\DLL$ and $\DTT$ of $\Lambda$ and $\overline{\Lambda}$ hyperons as a function of the hyperon transverse momentum $\pth$ up to 8 $\GeV$ in proton-proton collisions at $\sqrt s=200$\,$\GeV$ by the STAR experiment.
The new measurements have twice the hyperon statistics of previous publications in both the $\DLL$ and $\DTT$ cases. 
Our data are consistent with several model calculations within uncertainties, but one extreme scenario of polarized fragmentation functions for $D_{LL}$ assuming no flavor dependence is clearly disfavored.  
\vspace{4pt}

\par
We also report the first measurements of the spin transfer coefficients $\DLL$ and $\DTT$ for $\Lambda$ and $\overline{\Lambda}$ hyperons as a function of the fractional momentum $z$ of a jet carried by the hyperon with the same data sets, which provide direct probes of the corresponding polarized fragmentation functions.
Future measurements of spin transfer coefficients of hyperons in proton-proton collisions, in particular after the STAR forward detector upgrade at RHIC\,\cite{Brandenburg:2021hrv}, and in the DIS process at the Electron Ion Collider\,\cite{ABDULKHALEK2022122447}, will provide more information on the spin structure of the nucleon and the $\Lambda$ and $\overline{\Lambda}$ hyperons. 

\section*{ACKNOWLEDGMENTS}
We thank the RHIC Operations Group and RCF at BNL, the NERSC Center at LBNL, and the Open Science Grid consortium for providing resources and support.  This work was supported in part by the Office of Nuclear Physics within the U.S. DOE Office of Science, the U.S. National Science Foundation, National Natural Science Foundation of China, Chinese Academy of Science, the Ministry of Science and Technology of China and the Chinese Ministry of Education, the Higher Education Sprout Project by Ministry of Education at NCKU, the National Research Foundation of Korea, Czech Science Foundation and Ministry of Education, Youth and Sports of the Czech Republic, Hungarian National Research, Development and Innovation Office, New National Excellency Programme of the Hungarian Ministry of Human Capacities, Department of Atomic Energy and Department of Science and Technology of the Government of India, the National Science Centre and WUT ID-UB of Poland, the Ministry of Science, Education and Sports of the Republic of Croatia, German Bundesministerium f\"ur Bildung, Wissenschaft, Forschung and Technologie (BMBF), Helmholtz Association, Ministry of Education, Culture, Sports, Science, and Technology (MEXT), Japan Society for the Promotion of Science (JSPS) and Agencia Nacional de Investigaci\'on y Desarrollo (ANID) of Chile.
\bibliography{ref.bib}

\end{document}